%% file: fpcp2008_karliner.tex
\documentclass[twocolumn,twoside,slac_two]{revtex4}
\usepackage{graphicx}
\usepackage{fancyhdr}
\input color.tex

\def\beq{ \begin{equation} }
\def\eeq{ \end{equation} }
\def\mystrut{\vrule height 2.0ex depth 2.0ex width 0pt}
\def\tallstrut{\vrule height 4.0ex depth 2.5ex width 0pt}
\def\deepstrut{\vrule height 4.2ex depth 1.0ex width 0pt}
\def\medstrut{\vrule height 1ex depth 1.0ex width 0pt}
\def\lowstrut{\vrule height 0ex depth 0.9ex width 0pt }
\def\lowstrutA{\vrule height 0ex depth 1.1ex width 0pt }

\def\Tcc{\hbox{$\,{T^{\pm}\kern-2.0ex\lowstrut}_{\bar c c}\,$}}
\def\Tbb{\hbox{$\,{T^{\pm}\kern-2.0ex\lowstrutA}_{\bar b b}\,$}}

\begin{document}

\title{Heavy quark spectroscopy and prediction of bottom baryon masses}

%

\author{Marek Karliner}
\affiliation{Raymond and Beverly Sackler School of Physics and Astronomy, Tel Aviv University, Tel Aviv, Israel}

\begin{abstract}
I discuss several recent highly accurate
theoretical predictions for masses of baryons containing the $b$ quark, as well as
an effective supersymmetry between heavy quark baryons and mesons. I also suggest
some possibilities for observing exotic hadrons containing heavy quarks.
\end{abstract}

\maketitle

\thispagestyle{fancy}


\section{Introduction}
QCD describes hadrons as valence quarks in a sea of gluons and $\bar q q$ pairs.
At distances above $\sim 1~\hbox{GeV}^{-1}$ quarks acquire an effective
{\em constituent mass} due to chiral symmetry breaking.
A hadron can then be thought of as a bound state of constituent quarks.
In the simplest approximation the hadron mass $M$ is then given by the sum of the masses of
its constituent quarks $m_i$, first written down by
Sakharov and Zeldovich:
\beq
M = \sum_i m_i
\label{SakhZel}\eeq
the binding and kinetic energies are "swallowed" by the constituent quarks masses.

The first and most important correction comes from the color hyper-fine (HF)
chromo-magnetic interaction,
\begin{eqnarray}
M &=& \sum_i m_i + V_{i<j}^{HF(QCD)} \nonumber \\
\\
V^{HF(QCD)}_{ij} & = &
v_0 \,(\vec \lambda_i \cdot \vec \lambda_j)
\, {\vec \sigma_i \cdot \vec \sigma_j \over m_i m_j }
\, \langle \psi | \delta (r_i - r_j) | \psi \rangle
\nonumber
\label{HFQCD}
\end{eqnarray}
where $v_0$ gives the overall strength of the HF
interaction, $\vec \lambda_{i,j}$ are the $SU(3)$ color
matrices, $\sigma_{i,j}$ are the quark spin operators and
$|\psi \rangle$ is the hadron wave function.
This is a contact spin-spin interaction, analogous to the EM hyperfine interaction,
which is a product of the magnetic moments,
\beq
V^{HF(QED)}_{ij} \propto \vec \mu_i \cdot \vec \mu_j =
e^2 \, {\vec \sigma_i \cdot \vec \sigma_j \over m_i m_j }
\eeq
in QCD, the $SU(3)_c$ generators take place of the electric charge.

From eq.~(\ref{HFQCD}) many very accurate results have been obtained for
the masses of the ground-state hadrons. Nevertheless, several caveats are
in order. First, this is a low-energy phenomenological model, still awaiting
a rigorous derivation from QCD. It is far from providing a complete description
of the hadronic spectrum, but it provides excellent predictions for mass
splittings and magnetic moments.

The crucial assumptions of the model are:
\begin{itemize}
\item[(a)] HF interaction is considered as a perturbation
which does not change the wave function
\item[(b)]
effective masses of quarks are the same inside mesons and baryons
\item[(c)] there are no 3-body effects.
\end{itemize}
\section{Quark masses}
As the first example of the application of eq.~(\ref{HFQCD})
we can obtain the $m_c - m_s$
quark mass difference from the $\Lambda_c - \Lambda$
baryon mass difference:
\begin{eqnarray}
M(\Lambda_c) - M(\Lambda) =
\ \phantom{aaaaaaaaaaaaaaaaaaaaaaaa}
\nonumber \\
 = (m_u + m_d + m_c + V^{HF}_{ud} + V^{HF}_{uc} + V^{HF}_{dc})
\phantom{aaaaaa}\\
- \,\,(m_u + m_d + m_s + V^{HF}_{ud} + V^{HF}_{us} + V^{HF}_{ds})
\phantom{aaaaaa}
\nonumber\\
= m_c - m_s \phantom{aaaaaaaaaaaaaaaaaaaaaaaaaaaaaaaa}
\nonumber
\end{eqnarray}
where the light-quark HF interaction terms $V^{HF}_{ud}$ cancel between
the two expressions and the
HF interaction terms between the heavy and
light quarks vanish: $V^{HF}_{us} =  V^{HF}_{ds} = V^{HF}_{uc} = V^{HF}_{dc} = 0$,
since the $u$ and $d$ light quarks are coupled to a spin-zero diquark and
the HF interaction couples to the spin.

A second example shows how we can extract the ratio of the constituent quark masses
from the ratio of the the hyperfine splittings in the corresponding mesons.
The hyperfine splitting between $K^*$ and $K$ mesons is given by
\begin{eqnarray}
M(K^*) - M(K) = \phantom{aaaaaaaaaaaaaaaaaaaaaaaaa}
\nonumber \\
= v_0 \, { \vec \lambda_u \cdot \vec \lambda_s \over m_u m_s }
\left[ \left( \vec \sigma_u \cdot \vec \sigma_s\right)_{K^*}
-
\left( \vec \sigma_u \cdot \vec \sigma_s\right)_K \right]
\,\langle \psi | \delta(r) |\psi \rangle
\nonumber\\
=4 v_0 \, { \vec \lambda_u \cdot \vec \lambda_s \over m_u m_s }\,\langle \psi | \delta(r) |\psi \rangle
\phantom{aaaaaaa\,aaaaaaaaaaaaa}
\label{KHF}
\end{eqnarray}
and similarly for hyperfine splitting between $D^*$ and  $D$
with \,$s \rightarrow c$ \,everywhere.
From (\ref{KHF}) and its $D$ analogue we then immediately obtain
\begin{eqnarray}
{M(K^*) - M(K)  \over M(D^*) - M(D) } =
{
\tallstrut\displaystyle
4 v_0 \, { \vec \lambda_u \cdot \vec \lambda_s \over m_u m_s }\,\langle \psi | \delta(r) |\psi \rangle
\over
\deepstrut\displaystyle
4 v_0 \, { \vec \lambda_u \cdot \vec \lambda_c \over m_u m_c }\,\langle \psi | \delta(r) |\psi \rangle}
\approx {m_c \over m_s}
\end{eqnarray}

Table I shows the quark mass differences obtained from mesons and baryons \cite{KL2003}.
\clearpage
\begin{table}[t]
\centering
\caption{Quark mass differences from baryons and mesons}
\strut\kern-1em\includegraphics[width=87mm]{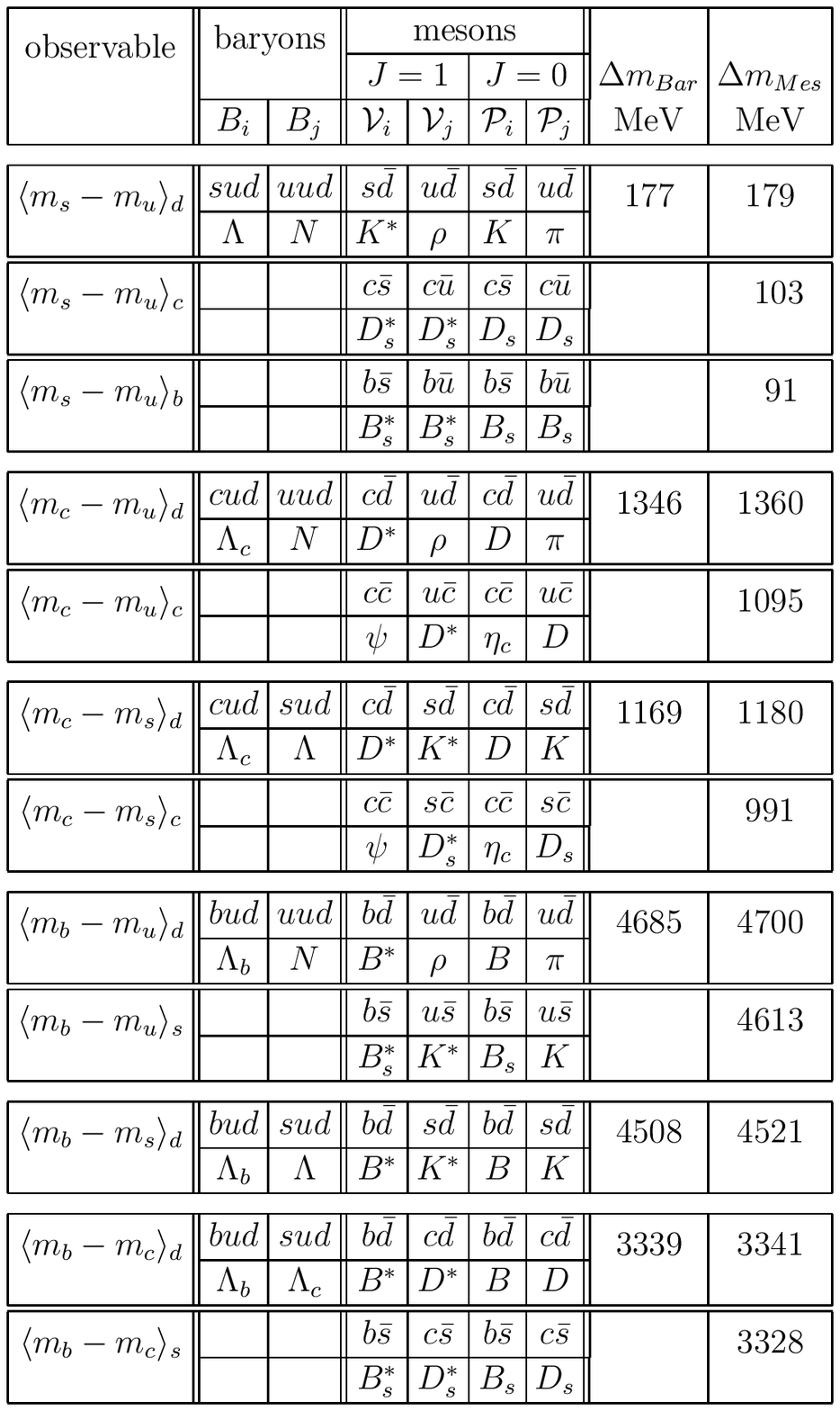}
\label{quark_mass_differences_table}
\end{table}
\strut\vskip-1cm
The mass difference between two quarks of different flavors denoted by $i$ and $j$ are seen
to have the same value to a good approximation when they are bound to a ``spectator"
quark of a given flavor.

On the other hand, Table I shows clearly that {\em constituent
quark mass differences depend strongly on the flavor of the spectator quark}. For example,
$m_s-m_d \approx 180$ MeV when the spectator is a light quark but the same mass difference
is only about 90 MeV when the spectator is a $b$ quark.

Since these are {\em effective masses},
we should not be surprised that their difference is affected by the environment, but the large size
of the shift is quite surprising and its quantitative derivation from QCD is an outstanding challenge for
theory.

\subsection{Color hyperfine splitting in baryons}

As an example of hyperfine splitting in baryons,
let us now discuss the HF splitting in the $\Sigma\ (uds)$ baryons.
$\Sigma^*$ has spin ${3\over2}$, so the $u$ and $d$ quarks must be in a state
of relative spin 1. The $\Sigma$ has isospin 1, so the wave function of
$u$ and $d$  is symmetric in flavor. It is also symmetric in space,
since in the ground state the quarks are in a relative $S$-wave.
On the other hand, the $u$-$d$ wave function is antisymmetric in color,
since the two quarks must couple to a {\bf 3}$^*$ of color to neutralize the
color of the third quark. The $u$-$d$ wave function must be antisymmetric
in \hbox{flavor\,$\times$\,spin$\,\times$\,space\,$\times$\,color,} so it follows it must be
symmetric in spin, i.e. $u$ and $d$ are coupled to spin one. Since $u$ and $d$ are in spin 1 state
in both $\Sigma^*$ and $\Sigma$ their HF interaction with each other cancels between the two
and thus the $u$-$d$ pair does not contribute to the \,$\Sigma^*-\Sigma$\, HF splitting,
\beq
M(\Sigma^*) - M(\Sigma) =
6 v_0 \, { \vec \lambda_u \cdot \vec \lambda_s \over m_u m_s }\,\langle \psi | \delta(r_{rs}) |\psi \rangle
\label{HFSigma}
\eeq
we can then use eqs. (\ref{KHF}) and (\ref{HFSigma}) to
compare the quark mass ratio obtained from mesons and baryons:

\begin{eqnarray}
\left({{m_c}\over{m_s}}\right)_{Bar} =
{{M_{\Sigma^*} - M_\Sigma}\over{M_{\Sigma_c^*} - M_{\Sigma_c}}} = 2.84
\nonumber\\
\\
\left({{m_c}\over{m_s}}\right)_{Mes} =
{{M_{K^*}-M_K}\over{M_{D^*}-M_D}}= 2.81
\nonumber\\
\nonumber
\end{eqnarray}

\begin{eqnarray}
 \left({{m_c}\over{m_u}}\right)_{Bar} =
{{M_\Delta - M_p}\over{M_{\Sigma_c^*} - M_{\Sigma_c}}} = 4.36
\nonumber\\
\\
\left({{m_c}\over{m_u}}\right)_{Mes} =
{{M_\rho-M_\pi}\over{M_{D^*}-M_D}}= 4.46
\nonumber\end{eqnarray}
We find the same value from mesons and baryons $\pm2\%$.

The presence of a fourth flavor gives us the possibility of obtaining a new
type of mass relation between mesons and baryons. The $\Sigma - \Lambda$ mass
difference is believed to be due to the difference between the $u-d$ and $u-s$
hyperfine interactions. Similarly, the $\Sigma_c - \Lambda_c$ mass
difference is believed to be due to the difference between the $u-d$ and $u-c$
hyperfine interactions. We therefore obtain the relation

\vbox{
\begin{eqnarray}
\left({
\displaystyle {1\over m_u^2} - {1\over m_u m_c}
\over
\displaystyle  {1\over m_u^2} - {1\over m_u m_s}}\right)_{\strut \kern-1ex Bar}
\kern-3.0ex
={{M_{\Sigma_c} - M_{\Lambda_c}}\over{M_{\Sigma} - M_\Lambda}}=2.16
\phantom{aaaaaaaaa}
\nonumber\\
\strut\kern-2em\\
\strut\kern-0.3em \left({
\displaystyle {1\over m_u^2} - {1\over m_u m_c}
\over
\displaystyle  {1\over m_u^2} - {1\over m_u m_s}}\right)_{\strut \kern-1exMes}
\kern-3.0ex
=
{{(M_\rho {-} M_\pi){-}(M_{D^*}{-}M_D)}
\over
{(M_\rho {-} M_\pi){-}(M_{K^*}{-}M_K)}}
=2.10\kern-1em\strut
\nonumber\end{eqnarray}
The meson and baryon relations agree to $\pm 3\%$.}

We can now
write down an analogous relation for hadrons containing the $b$
quark instead of the $s$ quark, obtaining the prediction
for splitting between $\Sigma_b$ and $\Lambda_b$ :

\begin{equation}
{{M_{\Sigma_b} - M_{\Lambda_b}}\over{M_{\Sigma} - M_\Lambda}} =
{{(M_\rho - M_\pi)-(M_{B^*}-M_B)}\over{(M_\rho - M_\pi)-(M_{K^*}-M_K)}}= 2.51
\label{Sigma_b_pred}
\end{equation}
yielding $M_{\Sigma_b} - M_{\Lambda_b} = 194 \,{\rm MeV}$ \cite{KL2003,Karliner:2006ny}.

\begin{figure}[h]
\centering
\includegraphics[width=60mm,angle=90]{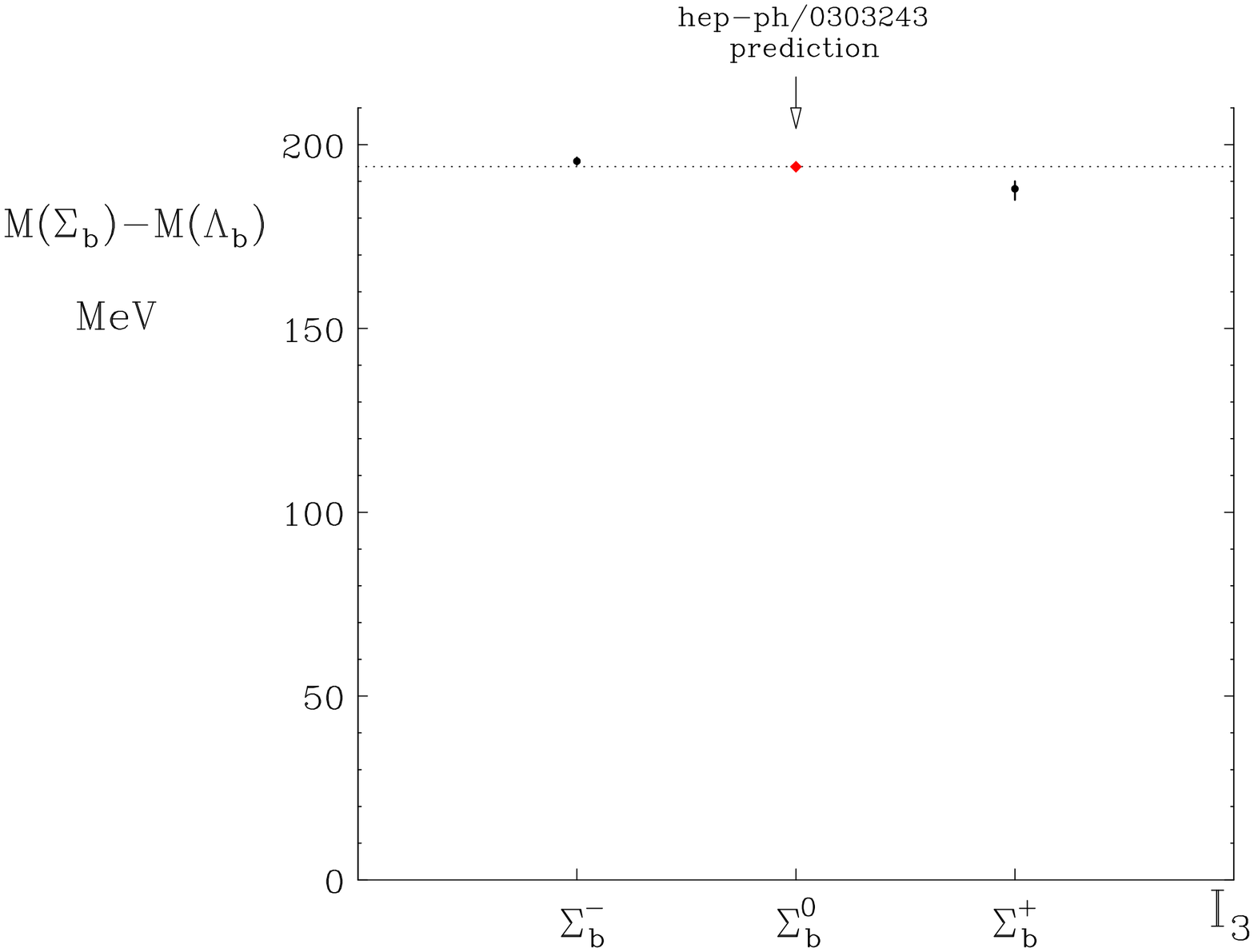}
\strut\kern3.1em\includegraphics[width=51mm,angle=90]{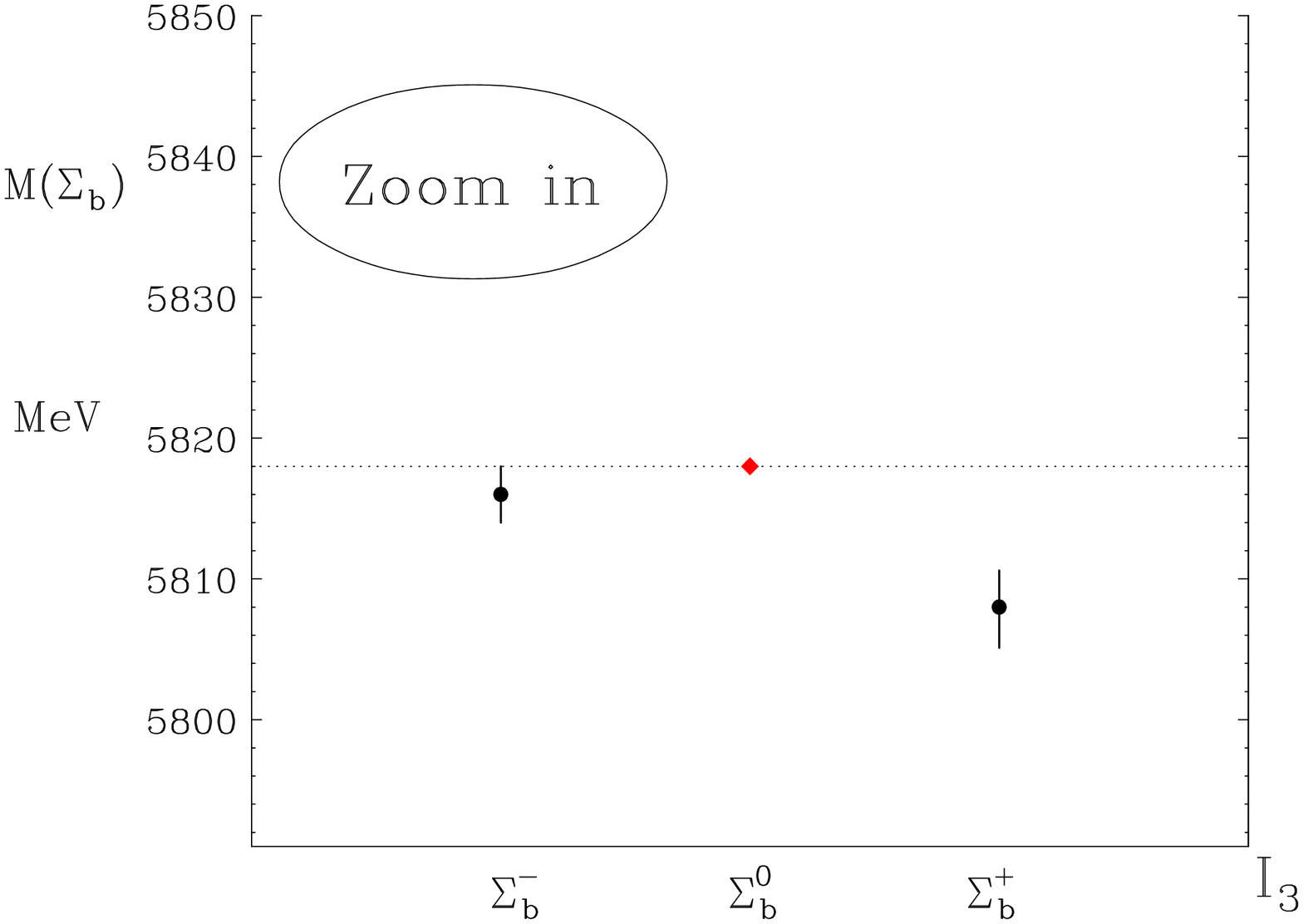}
\caption{Experimental results for $\Sigma_b^{\pm}$ masses
\cite{CDF_Sigma_b} vs the theoretical prediction in Ref.~\cite{KL2003}.}
\label{SigmabMasses}
\end{figure}

This splitting was recently measured by CDF \cite{CDF_Sigma_b}.
They obtained the masses of the $\Sigma_b^-$ and
$\Sigma_b^+$ from the decay \ $\Sigma_b \rightarrow \Lambda_b + \pi$ \ by
measuring the corresponding mass differences
\begin{eqnarray}
M(\Sigma_b^-) - M(\Lambda_b) = 195.5^{+1.0}_{{-}1.0}\,({\rm stat.}) \pm
0.1\, \hbox{(syst.) MeV}\kern-1em
\nonumber\\
\\
M(\Sigma_b^+) - M(\Lambda_b) = 188.0^{+2.0}_{{-}2.3}\,({\rm stat.}) \pm
0.1\, \hbox{(syst.) MeV}\kern-1em
\nonumber
\end{eqnarray}
with isospin-averaged mass difference
$M(\Sigma_b) - M(\Lambda_b) = 192$ MeV, as shown in Fig.~\ref{SigmabMasses}.
There is also the prediction for the spin splittings, good to 5\%
\begin{eqnarray}
M(\Sigma_b^*)- M(\Sigma_b) =\phantom{aaaaaaaaaaaaaaaaaaaaaaaa}
\nonumber\\
\\
=
{{M(B^*)- M(B)}\over{M(K^*)-M(K)}}\cdot [M(\Sigma^*)-M(\Sigma)]=
22 \,{\rm MeV}
\nonumber \end{eqnarray}
to be compared with 21  MeV from the
isospin-average of CDF measurements \cite{CDF_Sigma_b}.

The relation (\ref{Sigma_b_pred}) is based on the
assumption that the $qq$ and $q \bar q$ interactions have the same flavor
dependence. This automatically follows from the assumption that both
hyperfine interactions are inversely proportional to the products of the same
quark masses. But all that is needed here is the weaker assumption of same
flavor dependence,

\beq
{{V_{hyp}(q_i \bar q_j)}
\over{V_{hyp}(q_i \bar q_k)}}=
{{V_{hyp}(q_i q_j)}
\over{V_{hyp}(q_i q_k)}}
\end{equation}
This yields \cite{KL2003}
\begin{eqnarray}
\displaystyle
{{M_{\Sigma_b} - M_{\Lambda_b}}\over{(M_\rho - M_\pi)-(M_{B^*}-M_B)}} = 0.32
\approx
\nonumber\\
\approx \displaystyle
{{M_{\Sigma_c}{-}M_{\Lambda_c}}\over{(M_\rho {-} M_\pi){-}(M_{D^*}{-}M_D)}} = 0.33
\approx
\\
\displaystyle
\approx
{{M_{\Sigma}{-}M_\Lambda}\over{(M_\rho {-} M_\pi){-}(M_{K^*}{-}M_K)}} = 0.325
\phantom{aa}
\label{eq:newpred2}
\end{eqnarray}
The baryon-meson ratios are seen to be independent of the flavor $f$.

The challenge is to understand how and under what assumptions one can derive from
QCD the very simple model of hadronic structure at low energies which leads to such accurate
predictions.
\section{Effective Meson-Baryon \hbox{\strut\ \ \ \ Supersymmetry}}
Some of the results described above can be understood \cite{Karliner:2006ny}
by observing that in the hadronic spectrum there is an approximate
effective supersymmetry between
mesons and baryons related by replacing a light antiquark by a light
diquark.

\begin{figure}[t]
\centering
\strut\kern-1em\includegraphics[width=28mm,angle=270]{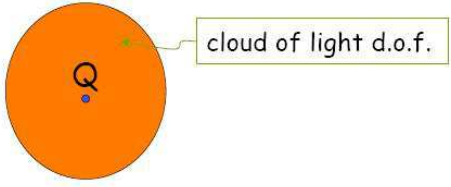}
\caption{A heavy quark coupled to ``brown muck" color antitriplet.}
\label{brown_muck}
\end{figure}

This supersymmetry transformation goes beyond the simple constituent quark
model. It assumes only a valence quark of flavor $i$ with a model independent
structure bound to
``light quark brown muck color antitriplet" of model-independent structure
carrying the quantum numbers of a light antiquark or a light diquark,
{\em cf.} Fig.~\ref{brown_muck}.
Since it assumes no model for the valence quark, nor the brown muck antitriplet
coupled to the valence quark,
it holds also for the quark-parton model in
which the valence is carried by a current quark and the rest of the hadron
is a complicated mixture of quarks and antiquarks.

\def\MM{{\cal M}}
\def\BB{{\cal B}}
\def\MUU{ {\cal V } }
\def\MUD{ {\cal P } }
\def\meddeepstrut{\vrule height 1.5ex depth 1.0ex width 1pt}

This light quark supersymmetry transformation, denoted here by $T^S_{LS}$,
connects a meson denoted by $\vert \MM(\bar q Q_i) \rangle$
and a baryon denoted by
$\vert \BB({[qq]\kern-0.5ex\phantom{\meddeepstrut}}_S Q_i) \rangle$
both containing the same valence quark
of some fixed flavor $Q_i$, $i=(u,s,c,b)$ and
a light color-antitriplet ``brown muck" state with
the flavor and baryon quantum numbers respectively of an antiquark $\bar
q$ ($u$ or $d$) and two light quarks coupled to a diquark of spin $S$.
\beq
T^S_{LS} \vert \MM(\bar q Q_i) \rangle
\quad \equiv \quad
\vert \BB({[qq]\kern-0.5ex\phantom{\meddeepstrut}}_S Q_i) \rangle
\label{numesbar}
\end{equation}
The mass difference between the meson and baryon related by this $T^S_{LS}$
transformation has been shown \cite{szmassqcd} to be independent of the quark
flavor $i$ for all four flavors $(u,s,c,b)$ when the contribution of the
hyperfine interaction energies is removed.
For the two cases of spin-zero\cite{szmassqcd} $S=0$ and
spin-one $S=1$ diquarks,
\begin{eqnarray}
M(N) - \,\tilde M\,(\,\rho\,)\, = 323~\rm{MeV} \approx
\nonumber\\
\approx M(\Lambda) - \tilde M(K^*) =321~\rm{MeV} \approx
\nonumber\\
\approx M(\Lambda_c) -  \tilde M(D^*) = 312~\rm{MeV} \approx
\nonumber\\
\approx M(\Lambda_b) - \tilde M(B^*)= 310~\rm{MeV} \phantom{aa\,\,}
\label{eq:mesbardif}
\end{eqnarray}

\begin{eqnarray}
\tilde M(\Delta) - \,\tilde M\,(\,\rho\,)\, = 517.56 ~\rm{MeV} \approx
\nonumber\\
\approx \tilde M(\Sigma) - \tilde M(K^*) =526.43 ~\rm{MeV} \approx
\nonumber\\
\approx \tilde M(\Sigma_c) -  \tilde M(D^*) = 523.95  ~\rm{MeV} \approx
\nonumber\\
\approx \tilde M(\Sigma_b) - \tilde M(B^*) = 512.45 ~\rm{MeV} \phantom{aa}\,\,
\label{eq:mesbardif2}
\end{eqnarray}

where
\beq
\tilde M(V_i)\equiv {{3M_{\MUU_i} + M_{\MUD_i} }\over {4}};
\label{tildev}
\end{equation}
are the weighted averages of vector and pseudoscalar meson masses,
denoted respectively by $M_{\MUU_i}$ and $M_{\MUD_i}$, which cancel their
hyperfine contribution,
and
\beq
\tilde M(\Sigma_i)\equiv {{2M_{\Sigma^*_i} + M_{\Sigma_i}}\over {3}} ;
\qquad
\tilde M(\Delta)\equiv{{2 M_\Delta +M_N}\over {3}}
\label{tildeb}
\end{equation}
are the analogous weighted averages
of baryon masses which cancel the hyperfine contribution between the diquark
and the additional quark.
\hfill\break
\strut
\vskip-1.0cm
\strut
\section{Magnetic Moments of Heavy Quark \hbox{\strut\ \ \ \ Baryons}}

In $\Lambda$, $\Lambda_c$ and $\Lambda_b$ baryons the light quarks are
coupled to
spin zero. Therefore the magnetic moments of these baryons are determined
by the
magnetic moments of the $s$, $c$ and $b$ quarks, respectively. The latter
are
proportional to the chromomagnetic moments which determine the hyperfine
splitting
in baryon spectra. We can use this fact to
predict the $\Lambda_c$ and $\Lambda_b$ baryon magnetic moments by
relating them to the
hyperfine splittings in the same way as given in the original
prediction \cite{DGG} of the $\Lambda$ magnetic moment,
\begin{eqnarray}
\label{maglam}
\mu_\Lambda=
-{\mu_p\over 3}\cdot {{M_{\Sigma^*} - M_\Sigma} \over{M_\Delta - M_N}}
&=&-0.61 \,{\rm n.m.}
\nonumber\\
\phantom{aaaaaaaaa}\qquad(\hbox{EXP}  &=&-0.61 \,{\rm n.m.})
\end{eqnarray}
\vskip-0.5cm
We obtain
\beq
\label{maglamc}
\mu_{\Lambda_c}= -2 {\mu_\Lambda}\cdot
{{M_{\Sigma_c^*} - M_{\Sigma_c}}\over{M_{\Sigma^*} - M_\Sigma}}
=\phantom{-}0.43\phantom{7}\,{\rm n.m.}
\end{equation}
\vskip-0.5cm
\beq
\label{maglamb}
\mu_{\Lambda_b}= \phantom{-} {\mu_\Lambda}\cdot
{{M_{\Sigma_b^*} - M_{\Sigma_b}} \over{M_{\Sigma^*} - M_{\Sigma}}}
=-0.067 \,{\rm n.m.}
\end{equation}
We hope these observables can be measured in foreseeable future and
view the predictions (\ref{maglamc}) and (\ref{maglamb}) as a challenge
for the experimental community.
\section{Testing Confining Potentials Through \hbox{\strut
\ \ \ \ \,Meson/Baryon HF Splitting Ratio}}
\begin{figure}[t]
\centering
\strut\kern-1em\includegraphics*[width=68mm]{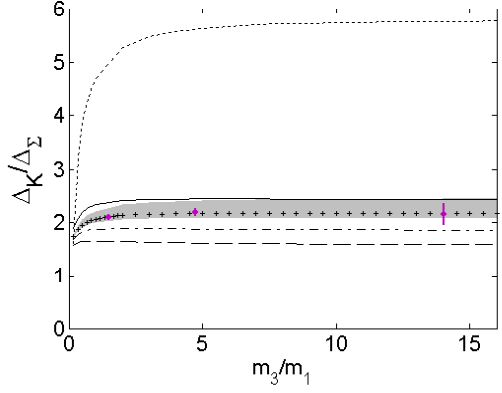}
\caption{Ratio of the hyperfine splittings in mesons and baryons, as function
of the quark mass ratio. Shaded region: Cornell potential for $0.2<k<0.5$;
crosses: Cornell, $k=0.28$; long dashes: harmonic oscillator; short dashes: Coulomb;
dot-dash: linear; continuous: logarithmic; thick dots: experimental data.}
\label{meson_baryon_ratio_fig}
\end{figure}

\vskip-0.1cm
\strut
The ratio of color hyperfine splitting in mesons and baryons
is a sensitive probe of the details of the confining
potential. This is because this ratio
depends only on the value of the wave function at the origin,
which in turn is
determined by the confining potential and by the ratio of quark
masses, as can be readily seen from eqs.~(\ref{KHF}) and (\ref{HFSigma}),
together with the fact that the color quark-antiquark interaction in mesons
is twice as strong as the quark-quark interaction in baryons,
$(\vec \lambda_u \cdot \vec \lambda_s)_{\hbox{\it \small meson}}
= 2 (\vec \lambda_u \cdot \vec \lambda_s)_{\hbox{\it \small baryon}}$.
We then have
\beq
{M_(K^*) - M(K) \over M(\Sigma^*) - M(\Sigma) } =
{4\over3}\, {  \langle \psi | \delta( \vec r_u - \vec r_s ) | \psi \rangle_{\hbox{\it \small meson}}
\over
  \langle \psi | \delta( \vec r_u - \vec r_s ) | \psi \rangle_{\hbox{\it \small baryon}} }
\label{MBHF}
\eeq
and analogous expressions with the $s$ quark replaced by another heavy quark $Q$.
From the experiment we have 3 data points for this ratio, with $Q=s,c,b$.
We can then compute the ratio (\ref{MBHF}) for 5 different representative confining
potentials and compare with experiment. The 5 potentials are
\begin{itemize}
\item harmonic oscillator
\item Coulomb interaction
\item linear potential
\item linear + Coulomb, i.e. Cornell potential
\item logarithmic
\end{itemize}
The results are shown in Fig.~\ref{meson_baryon_ratio_fig} and Table
\ref{meson_baryon_table} \cite{KerenZur:2007vp}.

\begin{table}[h]
\centering
\caption{Ratio of the hyperfine splittings in mesons and baryons, for
different potentials}
\strut\kern-0.5em\includegraphics[height=84mm,angle=270]{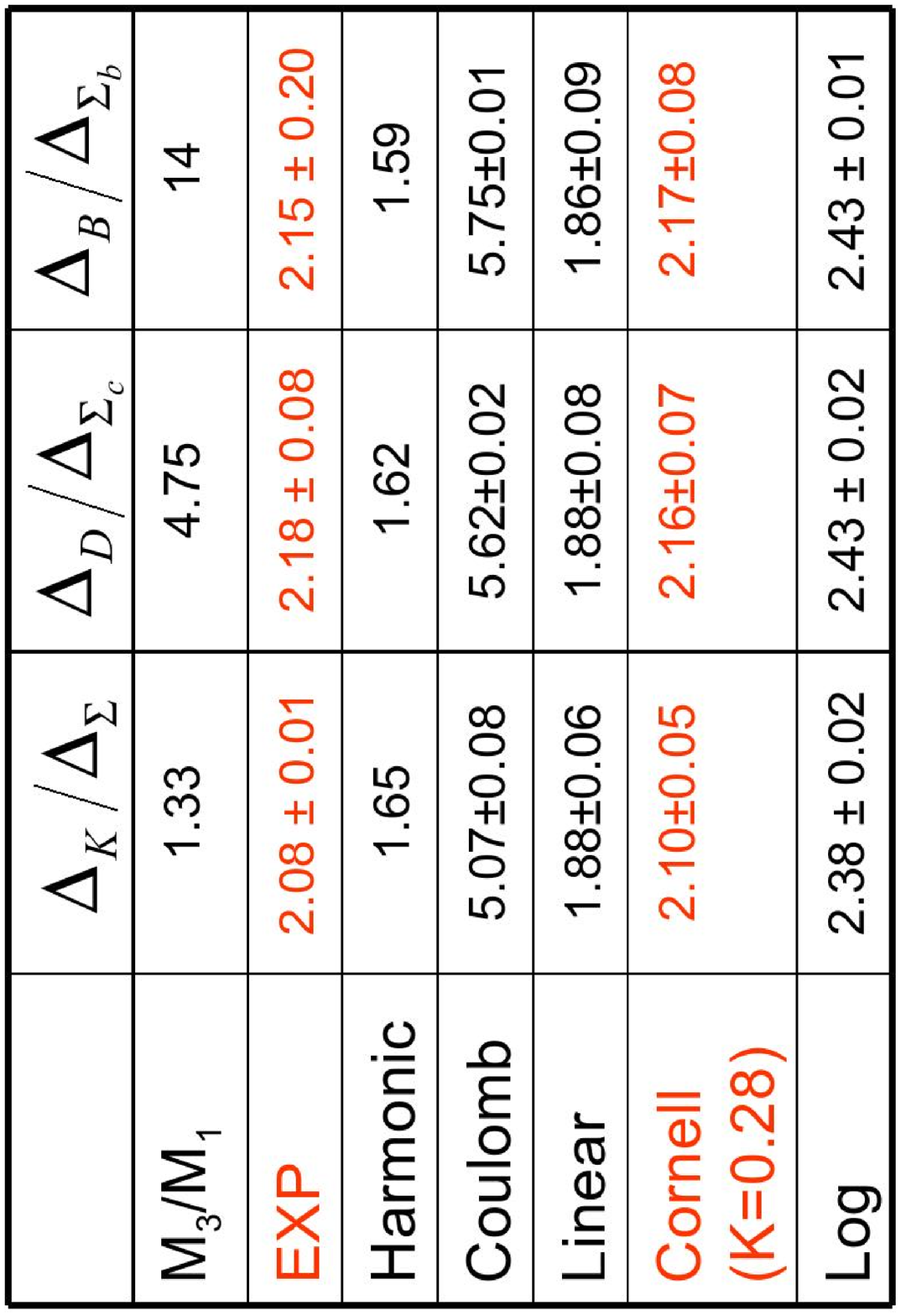}
\label{meson_baryon_table}
\end{table}

For all potentials which contain one coupling constant the coupling
strength cancels in the meson-baryon ratio. The Cornell potential
which is a combination of a Coulomb and linear potential contains
two couplings, one of which cancels in the meson-baryon ratio.
The remaining coupling is denoted by $k$.
The gray band corresponds to the range of values $0.2 < k < 0.5$
of the Cornell potential. The crosses correspond to $k=0.28$ which
is the value previously used to fit the charmonium data.
Clearly the Cornell potential with $k=0.28$
provides the best fit to the experiment.

\section{Predicting the Mass of \boldmath  $\Xi_Q$ \unboldmath  Baryons}

The $\Xi_Q$ baryons quark content is $Qsd$ or $Qsu$.
The name $\Xi_Q$ is a mnemonic for the fact that these states
can be obtained from ``ordinary" $\Xi$ ($ssd$ or $ssu$)
by replacing one of the $s$
quarks by a heavier quark $Q=c,b$. There is one important
difference, however.
In the ordinary $\Xi$, Fermi statistics
dictates that two $s$ quarks must couple to spin-1, while in the ground
state of $\Xi_Q$ the $(sd)$ and $(su)$ diquarks have spin zero.

Consequently, the $\Xi_b$ mass is given by the expression:
\begin{eqnarray}
\Xi_q=m_q+m_s+m_u-\frac{3v\langle \delta(r_{us}) \rangle}{m_um_s}
\end{eqnarray}
The $\Xi_b$ mass can thus be predicted using the known $\Xi_c$ baryon mass
as a
starting point and adding the corrections due to mass differences and HF
interactions:
\begin{eqnarray}
\Xi_b&=&\Xi_c + (m_b - m_c) +
\\
&-&\frac{3v}{m_um_s}\Bigg( \langle \delta(r_{us})
\rangle_{\Xi_b} -   \langle \delta(r_{us}) \rangle_{\Xi_c} \Bigg)
\end{eqnarray}

\subsection{Estimating \boldmath $(m_b - m_c)$ \unboldmath }
The mass difference $(m_b - m_c)$ can be obtained from experimental data using one
of the following
expressions:

\begin{itemize}

\item
We can simply take the difference of the masses of the $\Lambda_q$ baryons,
ignoring the differences in the HF interaction:
\begin{eqnarray}
m_b - m_c = \Lambda_b - \Lambda_c = 3333.2 \pm 1.2~.
\label{eq_lambda_b_lambda_c}
\end{eqnarray}

\item
We can use the spin averaged masses of the $\Lambda_q$ and $\Sigma_q$ baryons:
\begin{eqnarray}
m_b - m_c =
\phantom{aaaaaaaaaaaaaaaaaaaaaaa}
&&
\nonumber
\\
= \left(\frac{2 \Sigma_b^* + \Sigma_b+ \Lambda_b}{4}
- \frac{2\Sigma_c^* + \Sigma_c + \Lambda_c}{4}\right) = &&
\\
= 3330.4 \pm 1.8~.
\phantom{aaaaaaaaaaaaaaaaaaaa}
&&
\nonumber
\label{eq_sigma_b_sigma_c}
\end{eqnarray}

\item
Since the $\Xi_Q$ baryon contains a strange quark,
and the effective constituent quark masses depend on the spectator quark,
it might be better to use masses of mesons which contain both $s$ and
$Q$ quarks:
\begin{eqnarray}
m_b - m_c =
\phantom{aaaaaaaaaaaaaaaaaaaaaaa}
&&
\nonumber\\
= \left(\frac{3B_s^* + B_s}{4} - \frac{3D_s^* + D_s}{4}\right) =
\phantom{aaaa}
\\
 = 3324.6 \pm 1.4~.
\phantom{aaaaaaaaaaaaaaaaaaaa}
&&
\nonumber
\label{eq_B_s_D_s}
\end{eqnarray}

\end{itemize}

\subsection{\boldmath $\Xi_b$ Mass \unboldmath }
The corresponding results for $\Xi_b$ mass are summarized in Table \ref{tab_Xib}.

\begin{figure}[!b]
\centering
\strut\kern-1em\includegraphics[width=62mm,angle=270]{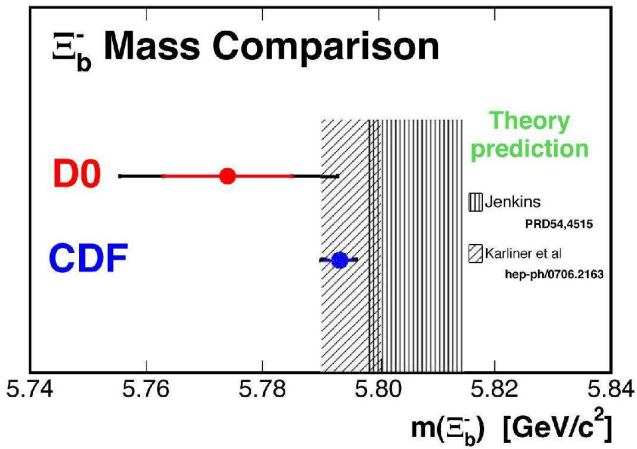}
\caption{Experimental results for the $\Xi_b$ mass compared with theoretical
predictions.}
\label{fig:Xib_EXP}
\end{figure}

\begin{table}[!htbp]
\caption{Predictions for the $\Xi_b$ mass with various confining
potentials and methods of obtaining the quark mass difference $m_b-m_c$ }
        \centering
\begin{tabular}{cccc} \hline \hline
\mystrut $m_b-m_c =$ & $\Lambda_b-\Lambda_c$
                     & ${\Sigma_b}-{\Sigma_c}$
                     & ${B_s}-{D_s}$          \\
& Eq.~(\ref{eq_lambda_b_lambda_c})
& Eq.~(\ref{eq_sigma_b_sigma_c})&eq.~(\ref{eq_B_s_D_s}) \\ \hline 
\mystrut No HF   correction & $5803\pm2$   & $5800\pm2$  & $5794\pm2$ \\
\mystrut Linear             & $5801\pm11$  & $5798\pm11$ & $5792\pm11$ \\
\mystrut Coulomb            & $5778\pm2$   & $5776\pm2$  & $5770\pm2$ \\
\mystrut Cornell            & $5799\pm7$   & $5796\pm7$  & $5790\pm7$ \\
\hline \hline
\end{tabular}
\label{tab_Xib}
\end{table}
On the basis of these results we predicted
\cite{Karliner:2007jp}
$M(\Xi_b) = 5795 \pm 5$~MeV. Our paper was submitted on June 14, 2007. The next day
CDF announced the result, following up on earlier but less precise D0 measurement.
These results are summarized in Fig.~\ref{fig:Xib_EXP}
and in Table \ref{tab:xib_exp}.

\begin{table}[!h]
\caption{Measurements of $\Xi_b$ mass at the
Fermilab Tevatron.  Errors on mass are (statistical, systematic).}
\label{tab:xib_exp}
\begin{center}
\begin{tabular}{c c c} \hline \hline
  & D0 \cite{Abazov:2007ub} & CDF \cite{Aaltonen:2007un} \\ \hline
Mass (MeV) & $5774 \pm 11 \pm 15$ & $5792.9 \pm 2.5 \pm 1.7$ \\
Width (MeV) & $37 \pm 8$ & $\sim 14$ \\
Significance & $5.5 \sigma$ & $7.8 \sigma$ \\ \hline \hline
\end{tabular}
\end{center}
\end{table}

\subsection{Mass of \boldmath $\Xi_b^\prime$ and $\Xi_b^*$ \unboldmath }
In the $\Xi_b^\prime$ baryon ($bsd$) the $(sd)$ diquark has $S=1$
and the total spin $= 1/2$.
In the $\Xi_b^*$ baryon ($bsd$) the $(sd)$ diquark also has $S=1$
and the total spin $= 3/2$.

The spin-averaged mass of these two states can be expressed as
\begin{eqnarray}
\frac{2\Xi_q^*+\Xi_q'}{3} = m_q + m_s + m_u
+ \frac{v\langle \delta(r_{us}) \rangle}{m_um_s}~,
\end{eqnarray}
and as for the $\Xi_b$ case, the following prediction can be given:
\begin{eqnarray}
\strut\kern-2em
\frac{2\Xi_b^*+\Xi_b'}{3}&=&\frac{2\Xi_c^*+\Xi_c'}{3} + (m_b - m_c) +
\nonumber\\
&+&
\frac{2\Xi_c^*+\Xi_c'-3\Xi_c}{12}
\left(\frac{ \langle \delta(r_{us}) \rangle_{\Xi_b}}
{\langle \delta(r_{us}) \rangle_{\Xi_c}}-1\right)
\end{eqnarray}
The predictions obtained using the same methods described above are given in
Table \ref{tab_Xib_star}.  Here the effect of the HF correction is negligible,
so the difference between the spin averaged mass $(2\Xi_b^*+\Xi_b')/3$ and
$\Xi_b$ is roughly $150-160$ MeV.

\begin{table}[h]
\caption{Predictions for the spin averaged $\Xi_b'$ and $\Xi_b^*$ masses
with various confining potentials and methods of obtaining the quark mass
difference $m_b-m_c$.
\label{tab_Xib_star}}
\begin{center}
\begin{tabular}{cccc} \hline \hline
$m_b-m_c =$ & $\Lambda_b-\Lambda_c$
            & ${\Sigma_b}-{\Sigma_c}$
            & ${B_s}-{D_s}$          \\
&Eq.~(\ref{eq_lambda_b_lambda_c})&Eq.~(\ref{eq_sigma_b_sigma_c})
&Eq.~(\ref{eq_B_s_D_s})\\ \hline 
No HF correction     & $5956\pm3$ & $5954\pm3$ & $5948\pm3$ \\
Linear               & $5957\pm4$ & $5954\pm4$ & $5948\pm4$ \\
Coulomb              & $5965\pm3$ & $5962\pm3$  & $5956\pm3$ \\
\red
Cornell              & $5958\pm3$ & $5955\pm3$ & $5949\pm3$
\black
\\ \hline \hline
\end{tabular}
\end{center}
\end{table}

The $\Xi_b^*-\Xi_b^\prime$ mass difference is more difficult to predict. It is
small, due to the large $m_b$:
\begin{equation}
\Xi_b^*-\Xi_b'=3v\Bigg(\frac{\langle \delta(r_{bs}) \rangle}{m_b m_s}
+\frac{\langle \delta(r_{bu}) \rangle}{m_b m_u}\Bigg)
\end{equation}

This expression is strongly dependent on the confinement model. In the results
given in Table \ref{tab_Xibstar_Xibprime} we have used $m_s/m_u=1.5\pm0.1$,
$m_b/m_c=2.95\pm0.2$.

\begin{table}
\caption{Predictions for $M(\Xi_b^*)-M(\Xi_b')$ with various confining
potentials.
\label{tab_Xibstar_Xibprime}}
\begin{center}
\begin{tabular}{cc} \hline \hline
                     & $\Xi_b^*-\Xi_b'$      \\ \hline 
No HF correction      & $24\pm2$  \\
Linear                & $28\pm6$ \\
Coulomb               & $36\pm7$  \\
\red Cornell               & $29\pm6$  \black \\ \hline \hline
\end{tabular}
\end{center}
\end{table}

In the context of $\Xi'_b$ and $\Xi_b^*$ masses it is worth mentioning
two elegant relations among bottom baryons \cite{Savage:1995dw} which
incorporate the effects of $SU(3)_f$ breaking:
\beq \label{eqn:sav1}
\Sigma_b + \Omega_b - 2 \Xi'_b   =  0~,
\eeq
\beq \label{eqn:sav2}
(\Sigma_b^* - \Sigma_b) + (\Omega_b^* - \Omega_b) - 2 (\Xi_b^* - \Xi'_b) =  0~,
\eeq
where isospin averaging is implicit.
\section{Predictions for Other \boldmath $b$ \unboldmath  Baryons}
Using methods similar to those discussed in the previous section
it is possible to make predictions for many other ground-state and excited
baryons containing the $b$ quark \cite{otherb}.

\subsection{\boldmath $\Omega_b$\unboldmath}
For the spin-averaged $\Omega_b$ mass we have
\begin{eqnarray}
\label{Omegab-spin-ave}
\frac{2M(\Omega_b^*)+M(\Omega_b)}{3}=
\phantom{aaaaaaaaaaaaaaaaaaaaaaaaaa}
\nonumber\\
=\frac{2M(\Omega_c^*)+M(\Omega_c)}{3}
+{(m_b-m_c)\medstrut}_{B_s-D_s}
\phantom{aaaaaaaaaa}
\\
 \nonumber
=6068.9\pm 2.4~\textnormal{MeV}
\phantom{\,\,aaaaaaaaaaaaaaaaaaaaaaaaaa}
\end{eqnarray}
For the HF splitting we obtain
\begin{eqnarray}
M(\Omega_b^*)-M(\Omega_b)=
\phantom{aaaaaaaaaaaaaaaaaaaaaaaaaa}
\nonumber\\
=
(M(\Omega_c^*)-M(\Omega_c))\frac{m_c}{m_b}\frac{\langle \delta(r_{bs})
\rangle_{\Omega_b}}{\langle \delta(r_{cs}) \rangle_{\Omega_c}}=
\phantom{aaaaaaaaaaa}
\\
=30.7 \pm1.3~\textnormal{MeV}
\phantom{aaaaaaaaaaaaaaaaaaaaaaaaaaaa}
\nonumber
\end{eqnarray}
leading to the following predictions:
\begin{equation}
\Omega_b^* = 6082.8\pm5.6~\rm{MeV}; ~ ~ ~ \Omega_b = 6052.1\pm5.6~\rm{MeV}
\label{omegab_pred}
\end{equation}

\subsection{Comparison with Other Approaches}
The sign in our prediction
\begin{equation}
M(\Sigma_b^*) - M(\Sigma_b) < M(\Omega_b^*) - M(\Omega_b)
\end {equation}
appears to be counterintuitive, since the color hyperfine
interaction  is inversely proportional to the quark mass. The expectation
value of the interaction with the same wave function for $\Sigma_b$ and
$\Omega_b$ violates our
inequality.  When wave function effects are included, the inequality
is still violated if the potential is linear, but is satisfied in
predictions which use the Cornell potential \cite{KerenZur:2007vp}.

This reversed inequality is  not predicted by other recent approaches
\cite{Ebert:2005xj,Roberts:2007ni,Jenkins:2007dm}  which all predict an
$\Omega_b$ splitting smaller than a $\Sigma_b$  splitting.

However the reversed inequality is also seen in the
corresponding charm experimental data,
\begin{eqnarray}
M(\Sigma_c^*) - M(\Sigma_c) & < & M(\Omega_c^*) - M(\Omega_c)
\nonumber
\\
64.3 \pm 0.5 \hbox{MeV}\strut\kern0.5em\strut  &&
\strut\kern0.2em\strut
 70.8\pm 1.5 \hbox{MeV}
\nonumber\\
\end{eqnarray}

This suggests that the sign of the $SU(3)$ symmetry breaking gives information
about the form of the potential. It is of interest to follow this clue
theoretically and experimentally.

\subsection{Additional States}
We have made predictions
\cite{Karliner:2007jp,otherb}
 for isospin splitting of $\Xi_b$,
and for orbital excitations of of $\Lambda_b$ and $\Xi_b$.
Our results are summarized in Table \ref{tab:comp}.

\begin{table*}
\caption{Comparison of predictions for $b$ baryons with those of some other
recent approaches \cite{Ebert:2005xj,Roberts:2007ni,Jenkins:2007dm} and with
experiment.  Masses quoted are isospin averages unless otherwise noted.
Our predictions are those based on the Cornell potential.
\label{tab:comp}}
\begin{center}
\begin{tabular}{cccccc} \hline \hline
         & \multicolumn{5}{c}{Value in MeV} \\
Quantity & Refs.\ \cite{Ebert:2005xj} & Ref.\ \cite{Roberts:2007ni} &
Ref.\ \cite{Jenkins:2007dm} & This work & Experiment \\ \hline
$M(\Lambda_b)$ & 5622 & 5612 & Input & Input & 5619.7$\pm$1.7 \\
$M(\Sigma_b)$ & 5805 & 5833 & Input & -- & 5811.5$\pm$2 \\
$M(\Sigma^*_b)$ & 5834 & 5858 & Input & -- & 5832.7$\pm$2 \\
$M(\Sigma^*_b) - M(\Sigma_b)$ & 29 & 25 & Input & 20.0$\pm$0.3
 & 21.2$^{+2.2}_{-2.1}$ \\
$M(\Xi_b)$ & 5812 & 5806$^a$ & Input & 5790--5800 & 5792.9$\pm$3.0$^b$ \\
$M(\Xi'_b)$  & 5937 & 5970$^a$ & 5929.7$\pm$4.4 & 5930$\pm$5 & -- \\
$\Delta M(\Xi^b)^c$ & -- & -- & -- & 6.4$\pm$1.6 & -- \\
$M(\Xi^*_b)$ & 5963 & 5980$^a$ & 5950.3$\pm$4.2 & 5959$\pm$4 & -- \\
$M(\Xi^*_b) - M(\Xi'_b)$ & 26 & 10$^a$ & 20.6$\pm$1.9 & 29$\pm$6 & -- \\
$M(\Omega_b)$ & 6065 & 6081 & 6039.1$\pm$8.3 & 6052.1$\pm$5.6 & --\\
$M(\Omega_b^*)$ & 6088 & 6102 & 6058.9$\pm$8.1 & 6082.8$\pm$5.6 & --\\
$M(\Omega_b^*) - M(\Omega_b)$ & 23 & 21 & 19.8$\pm$3.1 & 30.7$\pm$1.3 & -- \\
$M(\Lambda^*_{b[1/2]})$ & 5930 & 5939 & -- & $5929\phantom{.0} \pm 2$ & --\\
$M(\Lambda^*_{b[3/2]})$ & 5947 & 5941 & -- & $5940\phantom{.0} \pm 2$ & --\\
$M(\Xi^*_{b[1/2]})$ & 6119 & 6090 & -- & $6106\phantom{.0} \pm 4$ & --\\
$M(\Xi^*_{b[3/2]})$ & 6130 & 6093 & -- & $6115\phantom{.0} \pm 4$ & --\\
\hline \hline
\end{tabular}
\end{center}
\leftline{$^a$Value with configuration mixing taken into account; slightly
higher without mixing.}
\leftline{$^b$CDF \cite{Aaltonen:2007un} value of $M(\Xi_b^-)$.}
\leftline{$^c$$M$(state with $d$ quark) -- $M$(state with $u$ quark).}
\end{table*}

\section{\hbox{\boldmath $\Upsilon(5S){\rightarrow}\Upsilon(nS) \pi^+\pi^-$ and
Tetraquarks\unboldmath}}

The Belle Collaboration has recently reported \cite{BelleUpsilon5S}
anomalously large partial widths in
$\Upsilon(1S) \,\pi^+ \pi^-$
and
$\Upsilon(2S) \,\pi^+ \pi^-$
production at the $\Upsilon(5S)$,
more than two orders of magnitude larger than the corresponding
partial widths for $\Upsilon(4S)$, $\Upsilon(3S)$ or $\Upsilon(2S)$
decays.

We suggested \cite{Karliner:2008rc} that the large partial widths of
these channels might be due to their production by decays via an
intermediate
$\Tbb\pi^{\mp}$ state, where \Tbb\ denotes an isovector charged
 tetraquark $\bar b b u \bar d$ or $\bar b b \bar u d$,

\beq
\label{upstet}
\Upsilon(nS)\ \rightarrow \ \pi^{\mp} \Tbb \ \rightarrow \
\Upsilon(mS)\,\pi^-\pi^+
\eeq
The \Tbb\ might be an analogue of the
$Z(4430)$ resonance
reported by Belle in the summer of 2007 \cite{Z4430},
in the $\psi^\prime \pi^{\pm}$
invariant mass with
$M=4433 \pm 4\hbox{(stat)}\pm 2\hbox{(syst)}$ MeV and
\hbox{$\Gamma= 45{\,}^{+18}_{{-}13}\,\hbox{(stat)}
\, {}^{+30}_{{-}13}\,\hbox{(syst)}$ MeV.}
The $Z(4430)$ has not been seen in the $J/\psi \pi^\pm$ channel.
This report is awaiting
confirmation. If it is confirmed, it can be
a $\bar c c u \bar d$ tetraquark,
since it is an isovector and carries hidden charm.
Most calculations predict that such states are above the masses of two
separated heavy quark mesons as well as $Q \bar Q$ and
$u \bar d$ mesons. The $cu\bar c \bar d$ and  $bu\bar b\bar d$ states can
therefore decay into states like $D \bar D$ and $B \bar B$ as well as
$J/\psi \pi$ and $\Upsilon \pi$ with large widths. The narrow width
of the $Z(4430)$
and the lack of the $J/\psi \pi^\pm$ decay channel are therefore quite puzzling.
However a mechanism
\cite{mix} has been proposed in which the two color eigenstates can be mixed in
such a way that the otherwise dominant $D \bar D$ and $B \bar B$ are suppressed.
If confirmed, the existence of $Z(4430)$ would also make it very likely
that there is an analogous state in the bottom system.

Conservation of energy an momentum in the decay of $\Upsilon(5S)$ dictates
that
the scatter plot of $M_{inv}^2(\Upsilon(mS)\pi)$ vs. $E_{\pi}$ should be
linear, modulo
the width of $\Upsilon(5S)$ and $\Upsilon(mS)$. This can serve as a test of
the data
quality.  One should look for peaks in $M_{inv}(\Upsilon(mS)\pi)$, noting
that isospin
invariance dictates that the distribution of $M_{inv}(\Upsilon(mS)\pi^+)$ vs.
$E_{\pi^-}$
should be the same as $M_{inv}(\Upsilon(mS)\pi^-)$ vs. $E_{\pi^+}$, modulo
statistics.

\section{Open Questions}
Let me close with a list of current
open questions which I personally consider the most
interesting in this field
\begin{itemize}
\item
need to understand the XYZ states in the charm sector
   and their counterparts in the bottom sector, remembering that
  replacing charmed quark by bottom quark makes
   the binding stronger; this is an
  excellent challenge for experiment and theory

\item
  general question of exotics in QCD

\item
baryons with two heavy quarks:
  $\red cc \black q$, $\red bb \black q$, $\red bc \black q $
where $q=u,d$. So far the only positive
experimental report is about doubly charmed baryons from SELEX, but the very large
isospin splitting they reported between $ccu$ and $ccd$ is
very hard to understand.
\end{itemize}

\bigskip 
\begin{acknowledgments}
The work described in this talk was done in collaboration with
Boaz Keren-Zur, Harry Lipkin and Jon Rosner.
This research was supported in part by a grant from the
Israel Science Foundation administered by the Israel Academy of Sciences and
Humanities.
\end{acknowledgments}

\bigskip 

\end{document}

%% file: color.tex
\def\red{
}

\def\black{
}